\documentclass[aps,pra,showpacs,twocolumn,amsmath,amssymb]{revtex4}

\usepackage{graphicx}
\usepackage{dcolumn}
\usepackage{bm}

\begin{document}


\title{Entanglement between a Two-level System and a Quantum Harmonic Oscillator}

\author{Kwan-yuet Ho}%
 \email{kwyho@physics.umd.edu}
 \homepage{http://www.glue.umd.edu/~kwyho/}
\affiliation{%
Department of Physics, University of Maryland, College Park, MD 20742
}%

\date{January 2, 2008}

\begin{abstract}
The entanglement between a Pauli-like two-level system and a quantum harmonic oscillator enhanced by an interaction between them and a $\delta$-pulse sequence is studied, with the decoherence due to their coupling with a Markovian bath. Without the Markovian bath, the entanglement is enhanced to maximum possible values. With the Markovian bath, the entanglement is enhanced up to some time and then dissipated, with the system in thermal equilibrium with the Markovian bath after a very long time. The time for achieving the maximum entanglement shows discontinuous jumps over the parameters of decoherence.\end{abstract}

\maketitle

\section{\label{intro} Introduction}
Entanglement is a very important quantum characteristic that has no classical counterpart. Einstein, Podolsky and Rosen has proposed the famous EPR paradox in the context of two moving particles as a function of their coordinates and momenta, which are not commutative \cite{EPR}. Later, Bohm proposed the paradox in the context of two spins in terms of their spin in different directions \cite{bohm}. In these two examples of the illustration of entanglement, one is the entanglement between two continuous systems and another between two discrete systems. The entanglement between two discrete systems, especially two-level qubit like systems has been studied extensively with error-correction protocol. The entanglement between two continuous systems has been studied in depth in recent years \cite{braunstein}.

The entanglement between discrete systems and continuous systems is more complicated and it contains much richer behaviors. Examples include an atom (two-leveled) coupled to a bosonic field \cite{cmonroe}, or a system containing a superconducting charge qubit and a nanomechanical resonator \cite{schwab} \cite{tian}. The entanglement between two discrete systems and between two continuous systems can be exchanged \cite{bose_kim} with Jaynes-Cummings interaction \cite{jaynes_cummings}, and arbitrary entangled states between two continuous system can be generated arbitrarily with the help of a two-level system \cite{CKLaw}. On the other hand, a two-level system which are coupled to a bosonic dissipative environment are actually such a system. Sometimes an environment can be used to both enhance and dissipate the entanglement between another two systems to a certain extent \cite{cmonroe} \cite{aguado} \cite{CKLaw2}.

In this paper, the entanglement between a Pauli-like two-level system and a quantum harmonic oscillator is studied, while they are coupled to a common bosonic dissipative Markovian bath which induces decoherence. This will be organized as follows: in section \ref{model}, the model of such a whole system and the corresponding master equation are discussed. The different approaches of the quantification of entanglement are introduced in section \ref{quantify} and their advantages and disadvatanges are discussed. In section \ref{nodecoh}, without the presence of a Markovian bath and with the initial state being a Fock state $|0\rangle$ and a thermal state, the time-evolution of state is calculated and its entanglement is measured. In section \ref{decoh}, the presence of a Markovian bath is considered.

\section{\label{model} Model}
In this system, there are a Pauli-like two-level system, a quantum harmonic oscillator and a thermal Markovian bath. Therefore the free Hamiltonian of this system reads
\begin{equation}
\label{free_H} \mathcal{H}_0 = \hbar \omega_0 a^{\dag} a - \frac{\epsilon_z}{2} \sigma_z + \hbar \sum_{\mathbf{k}} \nu_{\mathbf{k}} {b_{\mathbf{k}}}^{\dag} b_{\mathbf{k}} .
\end{equation}
The first term is the energy of the oscillator. The bosonic operator $a$ (and $a^{\dag}$) is the annihilation (and creation) operator for this oscillator. The second term is the energy of the two-level system with difference between the two energy levels given by $\epsilon_z$. The third term is the energy of the bath with many degrees of freedom given by the momenta $\mathbf{k}$'s, with annihilation (and creation) operator being $b_{\mathbf{k}}$ (and ${b_{\mathbf{k}}}^{\dag}$).

The coupling between the two-level system and the oscillator is given by the following interaction \cite{jaynes_cummings}:
\begin{equation}
\label{JC_H} \mathcal{H}_{I} = - \frac{\lambda_0}{2} (a + a^{\dag}) \sigma_z ,
\end{equation}
where $\lambda_0$ is the coupling parameter. Such interaction can be found in many systems, such as a system consisting of a nanomechnical resonator and a single Cooper-pair box.

With only the interactions (\ref{JC_H}), the system would just oscillate between spin-up and spin-down state, as in Rabi oscillation. The entanglement would then oscillate between the lowest and highest values. In order to enhance the entanglement, a $\delta$-pulse sequence is added upon the two-level system as
\begin{equation}
\label{tunnel_H} \mathcal{H}_t = \frac{\pi}{2} \hbar \sum_n \delta\left(t- n \tau_0\right) \sigma_x ,
\end{equation}
for $n$ is any positive integers and $\tau_0 = \frac{\pi}{\omega_0}$. This $\delta$-pulse sequence gives the effects of putting $-i \sigma_x$ in the calculation, and it flips the spin of the two-level system. This delta function is an approximation of such flipping pulse and is valid is such a pulse is much stronger than the energy spacing of the quantum harmonic oscillator $\hbar \omega_0$ \cite{tian}.

Both the two-level system and the oscillator is coupled to a Markovian bath by the following interaction:
\begin{eqnarray}
\label{H_cbath} \mathcal{H}_{B} &=& \hbar \sum_{\mathbf{k}} g_{1 \mathbf{k}} ({b_{\mathbf{k}}}^{\dag} \sigma_{-} e^{-i (\nu_{\mathbf{k}} + \frac{\epsilon_z}{\hbar}) t} + \sigma_{+} b_{\mathbf{k}} e^{i (\nu_{\mathbf{k}} + \frac{\epsilon_z}{\hbar}) t} ) \\
\nonumber && + \hbar \sum_{\mathbf{k}} g_{2 \mathbf{k}} ({b_{\mathbf{k}}}^{\dag} a e^{-i (\nu_{\mathbf{k}} - \omega_0) t} + a^{\dag} b_{\mathbf{k}} e^{i (\nu_{\mathbf{k}} - \omega_0) t} ) ,
\end{eqnarray}
where $g_{1 \mathbf{k}}$ and $g_{2 \mathbf{k}}$ are coupling constants. These couplings give rise to the dissipative effects and decoherence.

The total Hamiltonian is given by
\begin{equation}
\label{total_H} \mathcal{H} = \mathcal{H}_0 + \mathcal{H}_{I} + \mathcal{H}_t + \mathcal{H}_B .
\end{equation}
In interaction picture, the Hamiltonian is given by
\begin{equation}
\label{H_int} V(t) = \exp\left(i \frac{\mathcal{H}_0}{\hbar} t\right) \mathcal{H} \exp\left(- i \frac{\mathcal{H}_0}{\hbar} t\right) .
\end{equation}
By calculation, $V(t)$ can be read as
\begin{equation}
\label{Vt} V(t) = \tilde{\mathcal{H}}_{JC} (t) + \tilde{\mathcal{H}}_t (t) + \tilde{\mathcal{H}}_B (t) ,
\end{equation}
where
\begin{eqnarray}
\label{int_HJC} \tilde{\mathcal{H}}_{I} (t) &=& - \frac{\lambda_0}{2} (a e^{-i \omega_0 t} + a^{\dag} e^{i \omega_0 t}) \sigma_z ,\\
\label{int_Ht} \tilde{\mathcal{H}}_t (t) &=& \frac{\pi}{2} \hbar \sum_n \delta(t-n \tau_0) (e^{i \frac{\epsilon_z}{\hbar} t} \sigma_{-} + e^{-i \frac{\epsilon_z}{\hbar} t} \sigma_{+}) ,\\
\nonumber \tilde{\mathcal{H}}_B (t) &=& \hbar \sum_{\mathbf{k}} g_{1 \mathbf{k}} ({b_{\mathbf{k}}}^{\dag} \sigma_{-} + \sigma_{+} b_{\mathbf{k}} ) \\
\label{int_HB} && + \hbar \sum_{\mathbf{k}} g_{2 \mathbf{k}} ({b_{\mathbf{k}}}^{\dag} a + a^{\dag} b_{\mathbf{k}} ) .
\end{eqnarray}

With the density matrix of the whole system in the interaction picture given by $\rho (t)$, its time evolution is given by the following dynamical equation
\begin{equation}
\label{dynamic_eqn} \frac{d \rho(t)}{d t} = \frac{1}{i \hbar} [V(t), \rho(t)] .
\end{equation}

However, the entanglement between the two-level system and the oscillator is the concern, while the thermal bath is considered because of its decoherence effect. To facilitate the analysis, we consider the reduced density matrix consisting only of the two-level system and the oscillator, while all the degrees of freedom of the bath are traced out, written mathematically as
\begin{equation}
\label{red_rho} \rho_s (t) = Tr_B (\rho (t)) .
\end{equation}
While tracing out, the bath term (\ref{int_HB}) is expanded up to second order. Suppose the bath is stationary and the dissipation does not depend on the memory, i.e., the dissipation is said to be Markovian. Markovian approximation is taken and finally a master equation is derived \cite{zubairy}. So the final equation is given as
\begin{widetext}
\begin{eqnarray}
\nonumber \frac{d \rho_s (t)}{dt} &=& i \alpha_0 \omega_0 \left[e^{-i \omega_0 t} (a \sigma_z \rho_s (t) - \rho_s (t) a \sigma_z) + e^{i \omega_0 t} (a^{\dag} \sigma_z \rho_s (t) - \rho_s (t) a^{\dag} \sigma_z)\right] \\
\nonumber && - \bar{n}_{\sigma} \frac{\Gamma}{2} (\sigma_{-} \sigma_{+} \rho_s(t) - 2 \sigma_{+} \rho_s(t) \sigma_{-} + \rho_s(t) \sigma_{-} \sigma_{+}) \\
\nonumber && - (\bar{n}_{\sigma}+1) \frac{\Gamma}{2} (\sigma_{+} \sigma_{-} \rho_s(t) - 2 \sigma_{-} \rho_s(t) \sigma_{+} + \rho_s(t) \sigma_{+} \sigma_{-}) \\
\nonumber && - \bar{n}_{r} \frac{\mathcal{C}}{2} (a a^{\dag} \rho_s(t) - 2 a^{\dag} \rho_s(t) a + \rho_s(t) a a^{\dag}) \\
\label{mastereqn} && - (\bar{n}_{r}+1) \frac{\mathcal{C}}{2} (a^{\dag} a \rho_s(t) - 2 a \rho_s(t) a^{\dag} + \rho_s(t) a^{\dag} a) \\
\nonumber && - i \frac{\pi}{2} \sum_n \delta\left(t- n \tau_0\right) \left[
\left(e^{i \frac{\epsilon_z}{\hbar} t} \sigma_{-} + e^{- i \frac{\epsilon_z}{\hbar} t} \sigma_{+} \right)
\rho_s(t)- \rho_s(t)  \left(e^{i \frac{\epsilon_z}{\hbar} t} \sigma_{-} + e^{- i \frac{\epsilon_z}{\hbar} t} \sigma_{+} \right)\right] ,
\end{eqnarray}
\end{widetext}
where
\begin{eqnarray}
\label{def_nsigma} \bar{n}_{\sigma} &=& \frac{1}{\exp\left(\frac{\epsilon_z}{k_B T}\right)-1} , \\
\label{def_nr} \bar{n}_{r} &=& \frac{1}{\exp\left(\frac{\hbar \omega_0}{k_B T}\right)-1} ,
\end{eqnarray}
and $\alpha_0$ is related to the coupling by
\begin{equation}
\label{def_alpha0} \alpha_0 = \frac{\lambda_0}{2 \hbar \omega_0} .
\end{equation}

With the master equation, the time evolution of the system consisting only the two-level system and the oscillator can be evaluated.

\section{\label{quantify} Quantifying Entanglement}
Entanglement is a quantum property that bears no classical analogue. For a system in its pure state, it is said to be entangled if and only if there are more than one terms in the Schmidt decomposition of the quantum state \cite{knight}, where the Schmidt decomposition is given by
\begin{equation}
\label{def_schmidt_dec} |\psi\rangle = \sum_n \sqrt{\lambda_n} |a_n\rangle |b_n\rangle ,
\end{equation}
where ${a_n}$'s and ${b_n}$'s form two orthogonal sets of kets in two systems. The quantification of the entanglement is often given by von Neumann entropy. While the entropy is not advantageous for computation, the participation ratio \cite{Lambert} is often used for the sake of computation. The inverse participation ratio is first used as a measure of localization of a wavefunction \cite{Thouless}, but it is extended to deal with the mixedness of a density matrix. It is defined as
\begin{equation}
\label{def_K} K = \frac{1}{\sum_n {\lambda_n}^2} = \frac{1}{Tr({\rho_r}^2)} ,
\end{equation}
which is the reciprocal of the purity for $\rho_r$ is the reduced density matrix for one of the subsystems. It is a reasonable measure because the subsystems of a pure entangled state is a mixed state. However, this runs into the problem when the system itself is in mixed state, giving rise to the situation that the subsystem becomes mixed even if the system is not entangled.

For a system in its mixed state, it is said to be entangled if and only if its density matrix can be expressed as a finite sum of tensor products of the subsystems:
\begin{equation}
\label{def_ent} \rho = \sum_i p_i (\rho_i^A \otimes \rho_i^B) .
\end{equation}
One of the ways to determine whether it is entangled is introduced by Peres \cite{peres}. It is proved that if the partial transpose of such a state consists of negative eigenvalues, then the state is entangled. The quantity negativity is a measure of the amount of these non-negative eigenvalues, defined mathematically as \cite{vidal_werner}
\begin{equation}
\label{def_negativity} \mathcal{N}(\rho) = \frac{||\rho^{PT}||_1-1}{2} ,
\end{equation}
where $\rho^{PT}$ denotes the partial transpose of $\rho$, and
$||A||_1$ denotes the tracenorm of the operator $A$, i.e.,
$||A||_1 = Tr(\sqrt{A A^{T}})$. A non-entangled state have a negativity equal to zero. However, the converse is not true except for composite systems of dimension $2 \times 2$ and $2 \times 3$ as shown by the Horodecki family \cite{horodecki}. It often happens that an entangled state has a negativity equals zero as well. Yet the negativity is still a good measure of entanglement for mixed states.

In this paper, both the participation ratio and negativity will be used as a measure to measure the entanglement.

\section{\label{nodecoh} Entanglement Without Decoherence}
Consider the system without the effect of decoherence, i.e., $\Gamma = 0$ and $\mathcal{C} = 0$. In that case, the entanglement between the two-level system and the oscillator will achieve a maximum eventually. The state can be analytically expressed at time $t = n \tau_0$ for $n$ being any positive integers, while can be numerically computed at all the other times.

\begin{figure}
 \scalebox{.5}{\includegraphics{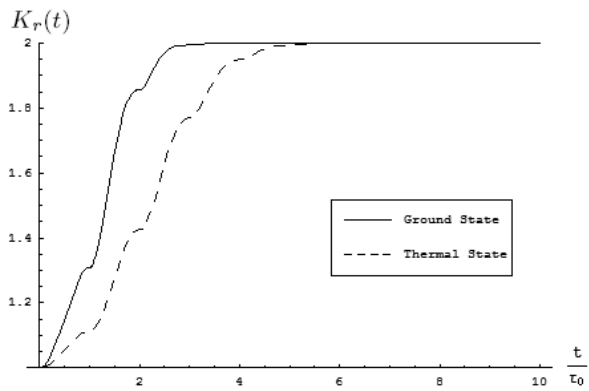}}
 \caption{\label{graph_K_nodecoh} Participation ratio of the oscillator $K_r (t)$ of the system of the two-level system and the oscillator with coupling strength $\lambda_0 = 0.2 \hbar \omega_0$, with the ground state and the thermal state ($\frac{\hbar \omega_0}{k_B T} = 0.74239$) as the initial states.}
 \end{figure}

\begin{figure}
 \scalebox{.5}{\includegraphics{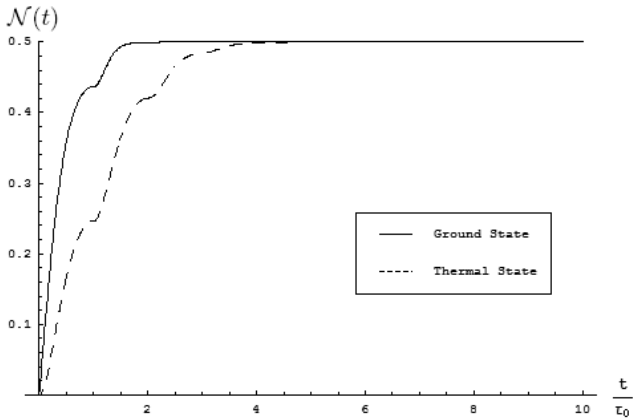}}
 \caption{\label{graph_negativity_nodecoh} Negativities $\mathcal{N} (t)$ of the system of the two-level system and the oscillator with coupling strength $\lambda_0 = 0.2 \hbar \omega_0$, with the ground state and the thermal state ($\frac{\hbar \omega_0}{k_B T} = 0.74239$) as the initial states.}
 \end{figure}

Suppose initially the state is a Fock state $|0\rangle$ given as
\begin{equation}
\label{initial_ground} |\psi(0)\rangle = \frac{1}{\sqrt{2}} (|\uparrow\rangle + |\downarrow\rangle) |0\rangle .
\end{equation}
Since it is a pure state, it can be solved using the Schrodinger's equation
\begin{equation}
\label{schreqn} [\tilde{H}_{I} (t) + \tilde{H}_{t} (t)] |\psi(t)\rangle = i \hbar \frac{\partial}{\partial t} |\psi(t)\rangle .
\end{equation}
And therefore, for $n\tau_0 \leq t < (n+1) \tau_0$,
\begin{eqnarray}
\label{pure_catstate}  |\psi(t)\rangle &=& \exp\left\{-\frac{i}{\hbar} \int_0^t dt' [\tilde{H}_{I} (t') + \tilde{H}_{t} (t')] \right\} |\psi(0)\rangle \\
\nonumber &=& \frac{1}{\sqrt{2}} \left(|\uparrow \rangle |- (-1)^n \tilde{\alpha} (t, n \tau_0) ] \rangle + |\downarrow\rangle | (-1)^n \tilde{\alpha} (t, n \tau_0) \rangle  \right) ,
\end{eqnarray}
where
\begin{equation}
\label{nodecoh_alphatilde} \tilde{\alpha} (t, n \tau_0) = 2 n \alpha_0 + \alpha_0 (1-e^{i \omega_0 (t-n\tau_0)}) ,
\end{equation}
and the coherent state is $|\alpha\rangle = D(\alpha) |0\rangle$ 
where $D(\alpha)$ is given by (\ref{disp_op}). After the interaction between the two systems, an obvious entanglement is established between the two subsystems, since the level of which state the two-level system is occupied is correlated to the signs of the coherent state. To measure its entanglement, its participation ratio and negativity are calculated. The participation ratios for both subsystems are
\begin{equation}
\label{K_nodecoh_grdstate} K = \frac{2}{1+\exp(-4 |\tilde{\alpha} (t, n \tau_0)|^2)} .
\end{equation}
And this is plotted as shown in Fig. \ref{graph_K_nodecoh}. The negativity has to be figured out by first finding the Schmidt decomposition of (\ref{pure_catstate}), which is
\begin{eqnarray}
\nonumber |\psi(t)\rangle &=& \sqrt{\frac{1+e^{-2 |\tilde{\alpha} (t, n \tau_0)|^2}}{2}} |+\rangle |f_+ (t)\rangle \\
\label{catstate_schmidt} && + \sqrt{\frac{1-e^{-2 |\tilde{\alpha} (t, n \tau_0)|^2}}{2}} |-\rangle |f_- (t)\rangle ,
\end{eqnarray}
where
\begin{eqnarray*}
|\pm\rangle &=& \frac{1}{\sqrt{2}} (|\uparrow\rangle \pm |\downarrow\rangle) , \\
|f_{\pm} (t)\rangle &=& \frac{1}{\sqrt{2 (1 \pm e^{-2 |\tilde{\alpha} (t, n \tau_0)|^2})}} \\
&& \cdot (|- (-1)^n \tilde{\alpha} (t, n \tau_0) \rangle \pm |(-1)^n \tilde{\alpha} (t, n \tau_0) \rangle ) .
\end{eqnarray*}
Vidal and Werner's analytic formula for negativity \cite{vidal_werner} is used to find the negativity of such system, which is
\begin{equation}
\label{neg_nodecoh_grdstate} \mathcal{N} = \frac{\sqrt{1-e^{-4 |\tilde{\alpha} (t, n \tau_0)|^2}}}{2} .
\end{equation}
The negativity (\ref{neg_nodecoh_grdstate}) is plotted as shown in Fig. \ref{graph_negativity_nodecoh}. When the system just starts to evolve for some small $t$, there are some overlapping between the two coherent states associated with $|\uparrow\rangle$ and $|\downarrow\rangle$. As time goes on, the coherent states gets further apart and $\langle - (-1)^n \tilde{\alpha} (t, n \tau_0) | (-1)^n \tilde{\alpha} (t, n \tau_0) ] \rangle \approx 0$, and thus system become more entangled. The negativity approaches its saturated (maximum) value of $\frac{1}{2}$ and the participation ratio $2$.

If the initial state is a thermal ground state at temperature $T$ given in terms of density matrix as
\begin{eqnarray}
\nonumber \rho_I(0) &=&
\left(\frac{|\uparrow\rangle+|\downarrow\rangle}{\sqrt{2}} \cdot
\frac{\langle\uparrow|+\langle\downarrow|}{\sqrt{2}}\right) \\
\nonumber && \otimes \left[(1-e^{-\frac{\hbar \omega_0}{k_B T}}) \sum_{r=0}^{\infty} e^{-\frac{ r \hbar \omega_0}{k_B T}} |r\rangle \langle r|\right] \\
\nonumber &=& \left(\frac{|\uparrow\rangle+|\downarrow\rangle}{\sqrt{2}} \cdot
\frac{\langle\uparrow|+\langle\downarrow|}{\sqrt{2}}\right) \\
\label{initial_rho_thermal} && \otimes \int d^2 \alpha \cdot P_0 (\alpha, \alpha^{*}) |\alpha\rangle \langle \alpha| ,
\end{eqnarray}
where $P_0 (\alpha, \alpha^{*})$ is the $P$-representation of the thermal state and is equal to
\begin{equation}
\label{P_thermal} P_0 (\alpha, \alpha^{*}) = \frac{1}{\pi \bar{n}_r} e^{-\frac{|\alpha|^2}{\bar{n}_r}} ,
\end{equation}
where the average boson distribution for the oscillator given by (\ref{def_nr}).
When $T = 0$, the state becomes the pure ground state $|0\rangle \langle 0|$ which can be again manipulated as shown previously for $\alpha = 0$. As in appendix \ref{ntau}, by (\ref{evol_rho_D}), the state can be calculated analytically at $t=n\tau_0$ as
\begin{eqnarray}
\nonumber \rho(n \tau_0) &=& |\uparrow \rangle \langle
\uparrow
| \otimes \int d^2 \alpha \cdot P_{\uparrow \uparrow} (\alpha,\alpha^{*}, n\tau_0) |\alpha \rangle \langle \alpha | \\
\nonumber &+& |\uparrow \rangle \langle \downarrow |
\otimes \int d^2 \alpha \cdot P_{\uparrow \downarrow} (\alpha,\alpha^{*}, n\tau_0) |\alpha
\rangle \langle \alpha | \\
\nonumber &+& |\downarrow \rangle \langle \uparrow |
\otimes \int d^2 \alpha \cdot P_{\downarrow \uparrow} (\alpha,\alpha^{*}, n\tau_0) |\alpha \rangle \langle \alpha | \\
\label{P_form_finalstate} &+& |\downarrow \rangle
\langle \downarrow | \otimes \int d^2 \alpha \cdot P_{\downarrow \downarrow}
(\alpha,\alpha^{*}, n\tau_0) |\alpha \rangle \langle \alpha | ,
\end{eqnarray}
where
\begin{eqnarray}
\label{form_P1} P_{\uparrow \uparrow} (\alpha,\alpha^{*}, n\tau_0) &=& \frac{1}{2 \pi \bar{n}_r}
e^{-\frac{|\alpha + (-1)^n 2 n \alpha_0|^2}{\bar{n}_r}}, \\
\nonumber P_{\uparrow \downarrow} (\alpha,\alpha^{*}, n\tau_0) &=& \frac{1}{2 \pi \bar{n}_r}
e^{\left(\frac{1}{\bar{n}_r}+2\right) 4 n^2 {\alpha_0}^2 } \cdot e^{-\frac{|\alpha|^2}{\bar{n}_r}}\\
\label{form_P2} &&  e^{2 (-1)^n
\left(\frac{1}{\bar{n}_r}+2\right) (\alpha-\alpha^{*}) n \alpha_0} ,\\
\nonumber P_{\downarrow \uparrow} (\alpha,\alpha^{*}, n\tau_0) &=& \frac{1}{2 \pi \bar{n}_r}
 e^{\left(\frac{1}{\bar{n}_r}+2\right) 4 n^2 {\alpha_0}^2 } e^{-e^{\frac{|\alpha|^2}{\bar{n}_r}}} \\
\label{form_P3} && \cdot 
e^{-2 (-1)^n \left(\frac{1}{\bar{n}_r}+2\right) (\alpha-\alpha^{*}) n \alpha_0} ,\\
\label{form_P4} P_{\downarrow \downarrow}
(\alpha,\alpha^{*}, n\tau_0) &=& \frac{1}{2 \pi \bar{n}_r}
e^{- \frac{|\alpha - (-1)^n 2 n \alpha_0|^2}{\bar{n}_r}}.
\end{eqnarray}
The participation ratios at $t = n \tau_0$ with respect to the two-level system and the oscillator are then given respectively by
\begin{eqnarray}
\label{K_thermal_sigma_ntau} K_{\sigma} (n \tau_0) &=& \frac{2}{1+\exp\left[-16 n^2 {\alpha_0}^2 (1+2 \bar{n}_r) \right]} ,\\
\label{K_thermal_res_ntau} K_{r} (n \tau_0) &=&  \frac{2 (1+2 \bar{n}_r)}{1+\exp\left[-\frac{16
n^2 {\alpha_0}^2 }{(1+2 \bar{n}_r)}\right]}.
\end{eqnarray}
As time goes on, $K_{\sigma} \rightarrow 2$ and $K_{r} \rightarrow 2 (1+2 \bar{n}_r)$. The discrepancy comes from the fact that the state is a mixed state.

The density matrix of the state can be computed at any time by solving the master equation (\ref{mastereqn}). This can be done numerically, but $P_{\uparrow \uparrow} (\alpha,\alpha^{*}, t)$ and $P_{\downarrow \downarrow} (\alpha,\alpha^{*}, t)$ can also be computed analytically and so is $K_{r}$, given by
\begin{equation}
\label{P_nodecoh_anytime} P_{\uparrow \uparrow / \downarrow \downarrow} (\alpha, \alpha^{*}, t)= \frac{1}{\pi \bar{n}_r} \exp\left[- \frac{|\alpha \mp (-1)^{(n+1)} \tilde{\alpha} (t,n\tau_0)|^2}{\bar{n}_r}\right],
\end{equation}
where $\tilde{\alpha} (t,n\tau_0)$ is defined in (\ref{nodecoh_alphatilde}). The reduced density matrix of the quantum oscillator side is
\begin{eqnarray}
\nonumber \rho_r(t) &=& \int d^2\alpha \cdot [P_{\uparrow\uparrow}(\alpha,\alpha^{*},t) + P_{\downarrow\downarrow}(\alpha,\alpha^{*},t)] |\alpha \rangle \langle \alpha| \\
\label{Ptotal} &=& \int d^2\alpha \cdot P(\alpha,\alpha^{*},t) |\alpha \rangle \langle \alpha| .
\end{eqnarray}
The participation ratio of the oscillator side can be evaluated directly by
\begin{eqnarray}
\nonumber \frac{1}{K_r} &=& \int d^2\alpha \int d^2\alpha' P(\alpha,\alpha^{*},t) P(\alpha',\alpha'^{*},t) \\
\label{eval_K} && \cdot e^{-|\alpha|^2-|\alpha'|^2+\alpha^{*} \alpha'+\alpha \alpha'^{*}} ,
\end{eqnarray}
and is found to be
\begin{equation}
\label{K_thermal_res_anytime} K_r (t) = \frac{2}{1+\exp\left[- \frac{4 |\tilde{\alpha} (t,n\tau_0)|^2}{2 \bar{n}_r + 1}\right]}.
\end{equation}
It is plotted as shown in figure \ref{graph_K_nodecoh}. Putting $t = n \tau_0$ in (\ref{K_thermal_res_anytime}) restores (\ref{K_thermal_res_ntau}), and putting $T = 0$ ($\bar{n}_r = 0$) restores the zero temperature case in (\ref{K_nodecoh_grdstate}).
With the numerical values of $\rho_s(t)$ computed, the negativity $\mathcal{N}(t)$, a better measure of entanglement, is calculated using (\ref{def_negativity}) and plotted as shown in Fig. \ref{graph_negativity_nodecoh}. It approaches the saturated value $\frac{1}{2}$ as time goes on, but at a slower rate than the case for the coherent state as the initial state.

\section{\label{decoh} Entanglement with Decoherence}
\begin{figure}
\scalebox{.5}{\includegraphics{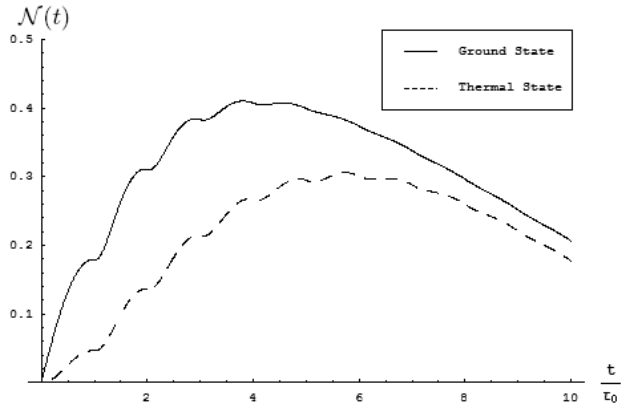}} \caption{\label{graph_negativity_decoh} Negativities $\mathcal{N} (t)$ of the system of the two-level system and the oscillator with coupling strength $\lambda_0 = 0.2 \hbar \omega_0$ with the ground state and the thermal state ($\frac{\hbar \omega_0}{k_B T} = 0.74239$) as the initial states with decoherence parameters $\Gamma = \mathcal{C} = \frac{0.01}{\tau_0}$.}
\end{figure}

Quantum systems are vulnerable to decoherence due to its coupling with the thermal bath stated in (\ref{H_cbath}), which is realistic since any quantum system cannot exist alone. To consider the the effect of decoherence, the state can be computed numerically by solving the master equations (\ref{mastereqn}) with all the terms. Its negativity can be computed numerically for every state by its definition (\ref{def_negativity}). After the interaction with the bath, the entanglement of the system increases until some point and it will eventually decrease, as shown in Fig. \ref{graph_negativity_decoh}. Although the negativity goes to zero at the end, it is still an open question whether the state is really non-entangled. Without decoherence, the state approaches the maximum negativity of $\frac{1}{2}$ after a long time. However, with even a small amount of coupling, the state reaches a maximum negativity which is less than $\frac{1}{2}$ at some certain time and then damps eventually. 

The state affected by decoherence can be studied by finding the Glauber-Sudershan representation of the reduced density matrix of the oscillator side analytically \cite{zubairy}. Its reduced density matrix can be found by tracing the spin components and be represented in Glauber-Sudershan $P$-representation as in (\ref{Ptotal}). 
Putting the expression for the representation to the master equation (\ref{mastereqn}), the Fokker-Planck equations for the oscillator state are derived (in ignorant of the last term of (\ref{mastereqn}))
\begin{widetext}
\begin{equation}
\label{fkeqn} \frac{\partial P_{\uparrow\uparrow / \downarrow\downarrow}}{\partial t} = \pm i \alpha_0 \omega_0 \left(e^{-i \omega_0 t} \frac{\partial P_{\uparrow\uparrow / \downarrow\downarrow}}{\partial \alpha^{*}} - e^{i \omega_0 t} \frac{\partial P_{\uparrow\uparrow / \downarrow\downarrow}}{\partial \alpha}\right) + \frac{\mathcal{C}}{2} \left(\frac{\partial}{\partial \alpha} \alpha + \frac{\partial}{\partial \alpha^{*}} \alpha^{*}\right) P_{\uparrow\uparrow / \downarrow\downarrow} + \mathcal{C} \bar{n}_{r} \frac{\partial^2 P_{\uparrow\uparrow / \downarrow\downarrow}}{\partial \alpha \partial \alpha^{*}} ,
\end{equation}
\end{widetext}
where the upper sign denote the term for $P_{\uparrow\uparrow}$ and the lower sign for $P_{\downarrow\downarrow}$. The last term of (\ref{mastereqn}) just switches the spin at time $t = n \tau_0$ where $\tau_0 = \frac{\pi}{\omega}$. Between these switches, the state can be fully described by the Fokker-Planck equations (\ref{fkeqn}). The solutions to these two equations can be found using the Green's function of the Fokker-Planck equation, as listed in the appendix \ref{fpeqn_green}. If the initial state is a ground state, then the Glauber-Sudarshan representation of the oscillator at any time $n \tau_0 \leq t < (n+1) \tau_0$ is given by
\begin{eqnarray}
\label{P_grdstate} && P_{\uparrow\uparrow / \downarrow\downarrow} (\alpha, \alpha^{*}, t)\\
\nonumber &=& \frac{1}{2 \pi \bar{n}_r (1-e^{-\mathcal{C} t})} \exp\left[-\frac{|\alpha \mp (-1)^{n+1} \tilde{\alpha}(t, n \tau_0)|^2}{\bar{n}_r (1-e^{-\mathcal{C} t})}\right] ,
\end{eqnarray}
and if the initial state is a thermal state, then it is
\begin{equation}
\label{P_thermalstate} P_{\uparrow\uparrow / \downarrow\downarrow} (\alpha, \alpha^{*}, t) = \frac{1}{2 \pi \bar{n}_r} \exp\left[-\frac{|\alpha \mp (-1)^{n+1} \tilde{\alpha}(t, n \tau_0)|^2}{\bar{n}_r }\right] .
\end{equation}
Here $\tilde{\alpha}(t, n \tau_0)$ denotes value of the modal coherent state in any time $t$ between $n \tau_0$ and $(n+1) \tau_0$, expressed as
\begin{eqnarray}
\label{decoh_alphatilde} \tilde{\alpha}(t, n \tau_0) &=& \frac{\alpha_0 \omega_0 e^{-\frac{\mathcal{C}}{2} (n-1) \tau_0}}{\omega_0 - \frac{i \mathcal{C}}{2}} \left\{n (1+e^{-\frac{\mathcal{C}}{2} \tau_0}) \right.\\
\nonumber && \left.+ e^{i \omega_0 (t - n \tau_0)} e^{-\frac{\mathcal{C}}{2} \tau_0} \left[e^{-(i \omega_0 + \frac{\mathcal{C}}{2}) (t-n \tau_0)} - 1\right]\right\} .
\end{eqnarray}
Note that this $\tilde{\alpha}(t, n \tau_0)$ is different from another one for non-decoherence case in (\ref{nodecoh_alphatilde}). When $\mathcal{C} \rightarrow 0$, $\tilde{\alpha}(n\tau_0, n\tau_0) \rightarrow 2 n \alpha_0$, which is the result found for the case without decoherence. However, for a non-zero $\mathcal{C}$, $\tilde{\alpha}(t, n \tau_0) \rightarrow 0$ for $t \rightarrow \infty$.

\begin{figure}
\scalebox{.5}{\includegraphics{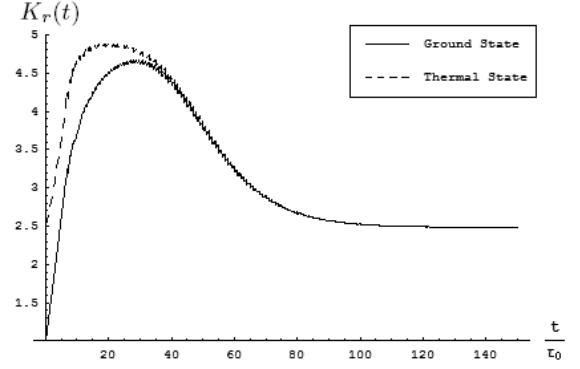}}
\caption{\label{graph_K_decoh} Participation ratio of the oscillator $K_r (t)$ of the system of the two-level system and the oscillator with coupling strength $\lambda_0 = 0.2 \hbar \omega_0$, with the ground state and the thermal state ($\frac{\hbar \omega_0}{k_B T} = 0.74239$) as the initial states. The decoherence parameter is $\mathcal{C} = 0.1$.}
\end{figure}

The representations in (\ref{P_grdstate}) and (\ref{P_thermalstate}) are important when studying the entanglement. By (\ref{eval_K}), the participation ratio for the ground state is,
\begin{equation}
\label{K_grdstate} K_r (t) = \frac{2 \left[2 \bar{n}_r (1-e^{-\mathcal{C} t})+1\right]}{1+\exp\left[-\frac{4 |\tilde{\alpha}(t,n \tau_0)|^2}{2 \bar{n}_r (1-e^{-\mathcal{C} t}) +1}\right]} ,
\end{equation}
and that for the thermal state is
\begin{equation}
\label{K_thermalstate} K_r (t) = \frac{2 \left[2 \bar{n}_r +1\right]}{1+\exp\left[-\frac{4 |\tilde{\alpha}(t,n \tau_0)|^2}{2 \bar{n}_r +1}\right]} .
\end{equation}
They are plotted as shown in Fig. \ref{graph_K_decoh}.
At the beginning, the values of $K$ are different since the ground state is pure at the beginning but the thermal state is already mixed. However, as time goes on, the two participation ratios coincides with each other. The reason is that the ground state is approaching the thermal state in order to achieve thermal equilibrium, as it can be seen in (\ref{P_grdstate}) that as $t \rightarrow \infty$, it is asymptotically approaching  (\ref{P_thermalstate}).

\begin{figure}
\scalebox{.5}{\includegraphics{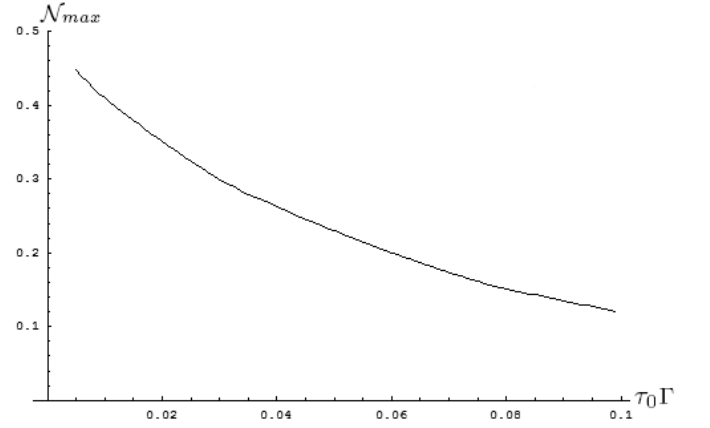}}
\caption{\label{graph_max_ent_coherent} Maximum negativity $\mathcal{N}_{max}$ of the system of the two-level system and the oscillator with coupling strength $\lambda_0 = 0.2 \hbar \omega_0$ with the coherent state as the initial states as a function of the decoherence parameters ( with $\Gamma = \mathcal{C}$) .}
\end{figure}

With ground state as the initial state, the maximum negativity as a function of the decoherence parameters (with $\Gamma = \mathcal{C}$) is plotted as shown in figure \ref{graph_max_ent_coherent}, which shows that the entanglement decreases as the decoherence strength increases. 
\begin{figure}
\scalebox{.5}{\includegraphics{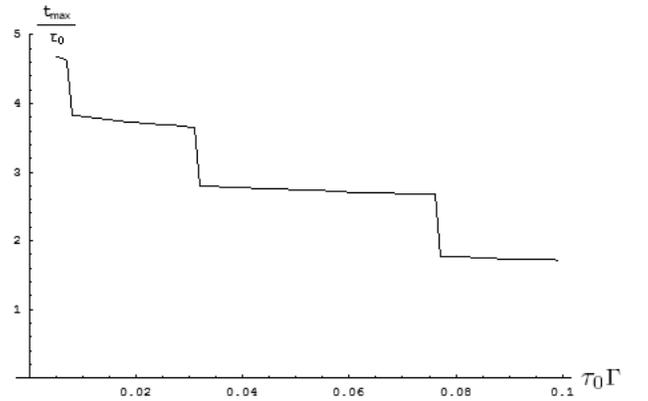}}
\caption{\label{graph_time_max_ent_coherent} Time $t_{max}$ (in terms of the unit of $\tau_0$) to achieve maximum negativity $\mathcal{N}_{max}$ of the system of the two-level system and the oscillator with coupling strength $\lambda_0 = 0.2 \hbar \omega_0$ with the coherent state as the initial states as a function of the decoherence parameters (with $\Gamma = \mathcal{C}$) .}
\end{figure}

\begin{figure}
\scalebox{0.5}{\includegraphics{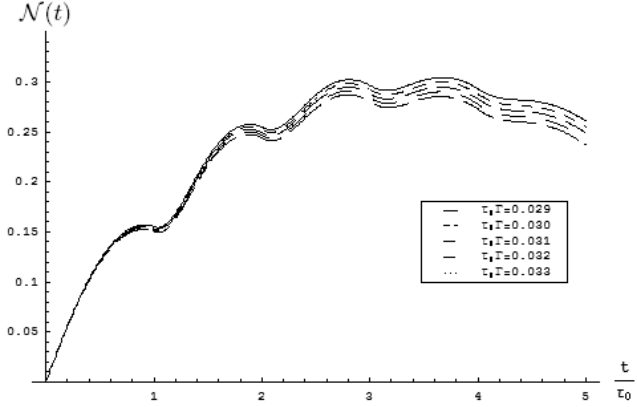}}
\caption{\label{graph_specific_delta_coherent} Negativities $\mathcal{N} (t)$ of the system of the two-level system and the oscillator with coupling strength $\lambda_0 = 0.2 \hbar \omega_0$ with the coherent state as the initial states with decoherence parameters $\Gamma = \mathcal{C}$ between $\frac{0.029}{\tau_0}$ and $\frac{0.034}{\tau_0}$.}
\end{figure}

\begin{figure}
\scalebox{0.45}{\includegraphics{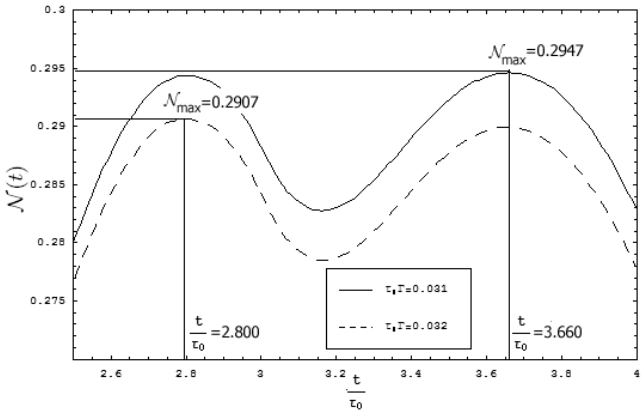}}
\caption{\label{graph_specific_delta_coherent_magnified} A magnified graph of the negativities $\mathcal{N} (t)$ of the system of the two-level system and the oscillator with coupling strength $\lambda_0 = 0.2 \hbar \omega_0$ with the coherent state as the initial states with decoherence parameters $\Gamma = \mathcal{C}$ being equal to $\frac{0.031}{\tau_0}$ and $\frac{0.032}{\tau_0}$.}
\end{figure}

It is worth to note that the time $t_{max}$ to achieve this maximum negativity decreases as the decoherence strength increases, with some jumps at a number of points as shown in figure \ref{graph_time_max_ent_coherent}. The jump can be explained by observing the time evolution of the entanglement with different values of $\Gamma (= \mathcal{C})$. As $\Gamma$ increases, the graph of the time evolution of the entanglement shifted downwards continuously as shown in figure \ref{graph_specific_delta_coherent}. Unlike the time evolution of the entanglement while there is no decoherence which increases monotonically (as in figure \ref{graph_negativity_nodecoh}), there are some ripples in the variation as time goes on when decoherence due to a thermal bath is present. The maximum negativity $\mathcal{N}_{max}$ still decreases continuously as $\Gamma$ increases, but the time $t_{max}$ to achieve this maximum get a jump because of these ripples, as shown in the magnified graph of the time variation of negativity in figure \ref{graph_specific_delta_coherent_magnified}.

\section{Conclusion}
It is shown that the Jaynes-Cummings interaction between a quantum oscillator and a Pauli-like two-level system, with the help of a delta jump in the two-level system, enhances the entanglement between the two systems to its maximum possible values, no matter the system is initially a Fock state $|0\rangle$ or a thermal state. The coupling of each subsystem to a Markovian bath causes dissipation of the entanglement. The system will be in thermal equilibrium with the Markovian bath after a very long time. The maximum time for achieving the maximum entanglement shows discontinuous jumps over the parameters of decoherence.

\begin{acknowledgments}
I wish to acknowledge Lin Tian for discussing this work and supplementing ideas.
\end{acknowledgments}

\appendix

\section{\label{ntau} Analytic Expression of the State Without Decoherence at $t=n\tau_0$}
Suppose the Dirac delta pulse (\ref{tunnel_H}) is imposed as the tunneling to the
system and there is no decoherence. The time evolution between the
pulses is given by \cite{tian}
\begin{equation}
\label{exp_U1} U_1 = D(\alpha_0 \sigma_z) e^{-i \pi a^{\dag} a}
D^{\dag}(\alpha_0 \sigma_z) ,
\end{equation}
where $D$ is the displacement operator given by
\begin{equation}
\label{disp_op} D(\alpha) = e^{\alpha a^{\dag}-\alpha^{*} a} .
\end{equation}
The effect of the Dirac delta pulse on the system is merely $(-i
\sigma_x)$, and thus the time evolution at time $t = n \tau_0$ is
given by
\begin{equation}
\label{exp_U} U(n \tau_0) = (-i \sigma_x U_1)^n .
\end{equation}
By $\sigma_x D(\alpha_0 \sigma_z) \sigma_x = D^{\dag}(\alpha_0
\sigma_z)$ and $e^{i \pi a^{\dag} a} D(\alpha_0 \sigma_z) e^{-i
\pi a^{\dag} a} = D^{\dag} (\alpha_0 \sigma_z)$, we have
\begin{equation}
\label{evol_n} U(n \tau_0) =
\left\{\begin{array}{ll}D^{\dag}(2 n \alpha_0 \sigma_z) & \mbox{ for even $n$} \\
i \sigma_x e^{-i \pi a^{\dag} a} D^{\dag}(2 n \alpha_0 \sigma_z) &
\mbox{ for odd $n$}\end{array} \right.  .
\end{equation}
As a result, if the system is a pure state desccribed by kets, the
initial state $|\psi_0\rangle$ will evolve to
\begin{eqnarray*}
\label{evol_ket_D} U(n \tau_0) |\psi_0\rangle .
\end{eqnarray*}
And if it is a mixed state, the state is described by density
matrices and at $t = n \tau_0$ the matrix is given by
\begin{equation}
\label{evol_rho_D} \rho(n \tau_0) =
\left\{\begin{array}{ll}D^{\dag}(2 n \alpha_0 \sigma_z) \rho_0
D(2 n \alpha_0 \sigma_z), \mbox{ for even $n$} \\
\sigma_x e^{-i \pi a^{\dag} a} D^{\dag}(2 n \alpha_0 \sigma_z)
\rho_0 D(2 n \alpha_0 \sigma_z) e^{i \pi a^{\dag} a} \sigma_x, \\
\mbox{ for odd $n$}\end{array} \right.   .
\end{equation}

\section{\label{fpeqn_green} Green's Function of the Fokker-Planck Equation}
The Fokker-Planck equations (\ref{fkeqn}) can be solved with the use of Green's function \cite{Louisell}. Suppose at time $t = n \tau_0$, the state is given by $P_{\uparrow \downarrow} (\alpha, \alpha^{*}, n \tau_0)$, the state at any time $n \tau_0 \leq t < (n+1) \tau_0$ is then given by
\begin{eqnarray}
&& P_{\uparrow\uparrow / \downarrow\downarrow} (\alpha, \alpha^{*}, n \tau_0 \leq t < (n+1) \tau_0)\\
\nonumber  &=& \int d^s\alpha \cdot P(\alpha, \alpha^{*},  t | \alpha', \alpha'^{*},  n \tau_0) P_{\uparrow \downarrow} (\alpha', \alpha'^{*}, n \tau_0) ,
\end{eqnarray}
where the Green's function is given by
\begin{eqnarray}
\nonumber P(\alpha, \alpha^{*},  t | \alpha', \alpha'^{*},  n \tau_0) = \frac{1}{\pi \bar{n}_r [1-e^{-\mathcal{C} (t-n\tau_0)}]} \\
\label{greenfcn} \cdot \exp\left[- \frac{|\alpha-\alpha' e^{-\frac{\mathcal{C}}{2} (t-n\tau_0)}+w(t,n \tau_0)|^2}{\bar{n}_r (1-e^{-\mathcal{C} t})}\right]  .
\end{eqnarray}
The function $w(t_2,t_1)$ is actually given by the driving force
\begin{eqnarray*}
w(t_2,t_1) &=& -i \int_{t_1}^{t_2} dt' (\mp \alpha_0 \omega_0) e^{i \omega_0 (t_2-t_1-t')} e^{-\frac{\mathcal{C}}{2} t'} \\
&=& \pm i \alpha_0 \omega_0 e^{i \omega_0 (t_2-t_1)} \frac{e^{-\left(i \omega_0+\frac{\mathcal{C}}{2}\right) t_2}-e^{-\left(i \omega_0+\frac{\mathcal{C}}{2}\right) t_1}}{-\left(i \omega_0+\frac{\mathcal{C}}{2}\right)} ,
\end{eqnarray*}
for the upper sign denotes the spin-up and the lower sign the spin-down. Then it can be proved that
\begin{equation}
\label{w_rel1} w(t, n\tau_0) = (-1)^n e^{-\frac{\mathcal{C}}{2} n \tau_0} w(t-n\tau_0,0) ,
\end{equation}
where
\begin{equation}
\label{w_rel2} w(t-n\tau_0,0) = \pm (-1)^n \alpha_0 \omega_0 e^{i \omega_0 t} \frac{e^{-(i \omega_0+\frac{\mathcal{C}}{2}) (t-n\tau_0)}-1}{\omega_0 - i \frac{\mathcal{C}}{2}} .
\end{equation}

\newpage 
\bibliography{entanglePRA}

\begin{thebibliography}{19}
\expandafter\ifx\csname natexlab\endcsname\relax\def\natexlab#1{#1}\fi
\expandafter\ifx\csname bibnamefont\endcsname\relax
  \def\bibnamefont#1{#1}\fi
\expandafter\ifx\csname bibfnamefont\endcsname\relax
  \def\bibfnamefont#1{#1}\fi
\expandafter\ifx\csname citenamefont\endcsname\relax
  \def\citenamefont#1{#1}\fi
\expandafter\ifx\csname url\endcsname\relax
  \def\url#1{\texttt{#1}}\fi
\expandafter\ifx\csname urlprefix\endcsname\relax\def\urlprefix{URL }\fi
\providecommand{\bibinfo}[2]{#2}
\providecommand{\eprint}[2][]{\url{#2}}

\bibitem[{\citenamefont{Einstein et~al.}(1935)\citenamefont{Einstein, Podolsky,
  and Rosen}}]{EPR}
\bibinfo{author}{\bibfnamefont{A.}~\bibnamefont{Einstein}},
  \bibinfo{author}{\bibfnamefont{B.}~\bibnamefont{Podolsky}}, \bibnamefont{and}
  \bibinfo{author}{\bibfnamefont{N.}~\bibnamefont{Rosen}},
  \bibinfo{journal}{Phys. Rev.} \textbf{\bibinfo{volume}{47}},
  \bibinfo{pages}{777} (\bibinfo{year}{1935}).

\bibitem[{\citenamefont{Bohm}(1989)}]{bohm}
\bibinfo{author}{\bibfnamefont{D.}~\bibnamefont{Bohm}},
  \emph{\bibinfo{title}{Quantum Theory}} (\bibinfo{publisher}{Prentice Hall},
  \bibinfo{address}{Englewoord Cliffs, NJ}, \bibinfo{year}{1989}).

\bibitem[{\citenamefont{Braunstein and van Loock}(2005)}]{braunstein}
\bibinfo{author}{\bibfnamefont{S.~L.} \bibnamefont{Braunstein}}
  \bibnamefont{and} \bibinfo{author}{\bibfnamefont{P.}~\bibnamefont{van
  Loock}}, \bibinfo{journal}{Rev. Mod. Phys.} \textbf{\bibinfo{volume}{77}},
  \bibinfo{pages}{513} (\bibinfo{year}{2005}).

\bibitem[{\citenamefont{Leibfried et~al.}(2003)\citenamefont{Leibfried, Blatt,
  Monroe, and Wineland}}]{cmonroe}
\bibinfo{author}{\bibfnamefont{D.}~\bibnamefont{Leibfried}},
  \bibinfo{author}{\bibfnamefont{R.}~\bibnamefont{Blatt}},
  \bibinfo{author}{\bibfnamefont{C.}~\bibnamefont{Monroe}}, \bibnamefont{and}
  \bibinfo{author}{\bibfnamefont{D.}~\bibnamefont{Wineland}},
  \bibinfo{journal}{Rev. Mod. Phys.} \textbf{\bibinfo{volume}{75}},
  \bibinfo{pages}{281} (\bibinfo{year}{2003}).

\bibitem[{\citenamefont{Irish et~al.}(2005)\citenamefont{Irish, Gea-Banacloche,
  Martin, and Schwab}}]{schwab}
\bibinfo{author}{\bibfnamefont{E.~K.} \bibnamefont{Irish}},
  \bibinfo{author}{\bibfnamefont{J.}~\bibnamefont{Gea-Banacloche}},
  \bibinfo{author}{\bibfnamefont{I.}~\bibnamefont{Martin}}, \bibnamefont{and}
  \bibinfo{author}{\bibfnamefont{K.~C.} \bibnamefont{Schwab}},
  \bibinfo{journal}{Phys. Rev. B} \textbf{\bibinfo{volume}{72}},
  \bibinfo{pages}{195410} (\bibinfo{year}{2005}).

\bibitem[{\citenamefont{Tian}(2005)}]{tian}
\bibinfo{author}{\bibfnamefont{L.}~\bibnamefont{Tian}}, \bibinfo{journal}{Phys.
  Rev. B} \textbf{\bibinfo{volume}{72}}, \bibinfo{pages}{195411}
  (\bibinfo{year}{2005}).

\bibitem[{\citenamefont{Lee et~al.}(2006)\citenamefont{Lee, Paternostro, Kim,
  and Bose}}]{bose_kim}
\bibinfo{author}{\bibfnamefont{J.}~\bibnamefont{Lee}},
  \bibinfo{author}{\bibfnamefont{M.}~\bibnamefont{Paternostro}},
  \bibinfo{author}{\bibfnamefont{M.~S.} \bibnamefont{Kim}}, \bibnamefont{and}
  \bibinfo{author}{\bibfnamefont{S.}~\bibnamefont{Bose}},
  \bibinfo{journal}{Phys. Rev. Lett.} \textbf{\bibinfo{volume}{96}},
  \bibinfo{pages}{080501} (\bibinfo{year}{2006}).

\bibitem[{\citenamefont{Jaynes and Cummings}(1963)}]{jaynes_cummings}
\bibinfo{author}{\bibfnamefont{E.~T.} \bibnamefont{Jaynes}} \bibnamefont{and}
  \bibinfo{author}{\bibfnamefont{F.~W.} \bibnamefont{Cummings}},
  \bibinfo{journal}{Proc. IEEE} \textbf{\bibinfo{volume}{51}},
  \bibinfo{pages}{89} (\bibinfo{year}{1963}).

\bibitem[{\citenamefont{Law and Eberly}(1996)}]{CKLaw}
\bibinfo{author}{\bibfnamefont{C.~K.} \bibnamefont{Law}} \bibnamefont{and}
  \bibinfo{author}{\bibfnamefont{J.~H.} \bibnamefont{Eberly}},
  \bibinfo{journal}{Phys. Rev. Lett.} \textbf{\bibinfo{volume}{76}},
  \bibinfo{pages}{1055} (\bibinfo{year}{1996}).

\bibitem[{\citenamefont{Contreras-Pulido and Aguado}(2007)}]{aguado}
\bibinfo{author}{\bibfnamefont{L.~D.} \bibnamefont{Contreras-Pulido}}
  \bibnamefont{and} \bibinfo{author}{\bibfnamefont{R.}~\bibnamefont{Aguado}}
  (\bibinfo{year}{2007}), \eprint{arXiv:0710.3576v1}.

\bibitem[{\citenamefont{Poon and Law}(2007)}]{CKLaw2}
\bibinfo{author}{\bibfnamefont{P.~S.~Y.} \bibnamefont{Poon}} \bibnamefont{and}
  \bibinfo{author}{\bibfnamefont{C.~K.} \bibnamefont{Law}},
  \bibinfo{journal}{Phys. Rev. A} \textbf{\bibinfo{volume}{76}},
  \bibinfo{pages}{012333} (\bibinfo{year}{2007}).

\bibitem[{\citenamefont{Scully and Zubairy}(1997)}]{zubairy}
\bibinfo{author}{\bibfnamefont{M.~O.} \bibnamefont{Scully}} \bibnamefont{and}
  \bibinfo{author}{\bibfnamefont{M.~S.} \bibnamefont{Zubairy}},
  \emph{\bibinfo{title}{Quantum Optics}} (\bibinfo{publisher}{Cambridge},
  \bibinfo{address}{Cambridge, England}, \bibinfo{year}{1997}).

\bibitem[{\citenamefont{Ekert and Knight}(1995)}]{knight}
\bibinfo{author}{\bibfnamefont{A.}~\bibnamefont{Ekert}} \bibnamefont{and}
  \bibinfo{author}{\bibfnamefont{P.~L.} \bibnamefont{Knight}},
  \bibinfo{journal}{Am. J. Phys.} \textbf{\bibinfo{volume}{63}},
  \bibinfo{pages}{5} (\bibinfo{year}{1995}).

\bibitem[{\citenamefont{Lambert et~al.}(2005)\citenamefont{Lambert, Emary, and
  Brandes}}]{Lambert}
\bibinfo{author}{\bibfnamefont{N.}~\bibnamefont{Lambert}},
  \bibinfo{author}{\bibfnamefont{C.}~\bibnamefont{Emary}}, \bibnamefont{and}
  \bibinfo{author}{\bibfnamefont{T.}~\bibnamefont{Brandes}},
  \bibinfo{journal}{Phys. Rev. A} \textbf{\bibinfo{volume}{71}},
  \bibinfo{pages}{053804} (\bibinfo{year}{2005}).

\bibitem[{\citenamefont{Edwards and Thouless}(1972)}]{Thouless}
\bibinfo{author}{\bibfnamefont{J.~T.} \bibnamefont{Edwards}} \bibnamefont{and}
  \bibinfo{author}{\bibfnamefont{D.~J.} \bibnamefont{Thouless}},
  \bibinfo{journal}{J. Phys. C} \textbf{\bibinfo{volume}{5}},
  \bibinfo{pages}{807} (\bibinfo{year}{1972}).

\bibitem[{\citenamefont{Peres}(1996)}]{peres}
\bibinfo{author}{\bibfnamefont{A.}~\bibnamefont{Peres}},
  \bibinfo{journal}{Phys. Rev. Lett.} \textbf{\bibinfo{volume}{77}},
  \bibinfo{pages}{1413} (\bibinfo{year}{1996}).

\bibitem[{\citenamefont{Vidal and Werner}(2002)}]{vidal_werner}
\bibinfo{author}{\bibfnamefont{G.}~\bibnamefont{Vidal}} \bibnamefont{and}
  \bibinfo{author}{\bibfnamefont{R.~F.} \bibnamefont{Werner}},
  \bibinfo{journal}{Phys. Rev. A} \textbf{\bibinfo{volume}{65}},
  \bibinfo{pages}{032314} (\bibinfo{year}{2002}).

\bibitem[{\citenamefont{Horodecki}(1997)}]{horodecki}
\bibinfo{author}{\bibfnamefont{P.}~\bibnamefont{Horodecki}}
  (\bibinfo{year}{1997}), \eprint{quant-ph/9703004}.

\bibitem[{\citenamefont{Louisell}(1973)}]{Louisell}
\bibinfo{author}{\bibfnamefont{W.~H.} \bibnamefont{Louisell}},
  \emph{\bibinfo{title}{Quantum Statistical Properties of Radiation}}
  (\bibinfo{publisher}{John Wiley and Sons}, \bibinfo{year}{1973}).

\end{thebibliography}

\end{document}